\documentclass[aps,prl,reprint,superscriptaddress,showpacs]{revtex4-1}
\usepackage{xcolor}
\usepackage{color}
\usepackage{amsmath}
\usepackage{amssymb}
\usepackage{wasysym}
\usepackage{mathcomp}
\usepackage{graphicx}
\usepackage{dcolumn}
\usepackage{bm}
\usepackage{hyperref}
\usepackage[mathlines]{lineno}
\usepackage{comment}
\includecomment{pablocomments}
\usepackage{tikz}
\usepackage{pgfplots}
\usetikzlibrary{shapes}

\newcommand*{\tikzbullet}[2]{%
  \setbox0=\hbox{\strut}%
  \begin{tikzpicture}
    \filldraw[draw=#1,fill=#2] (0,3\ht0) circle[radius=.25em];
  \end{tikzpicture}%
}

\newcommand*{\tikzrectangle}[2]{%
  \setbox0=\hbox{\strut}%
  \begin{tikzpicture}
    \filldraw[draw=#1,fill=#2] (0,3\ht0) rectangle ++(5pt,5pt);
  \end{tikzpicture}%
}

\newcommand*{\tikztriangle}[2]{%
  \setbox0=\hbox{\strut}%
  \begin{tikzpicture}
    \node[draw=#1,fill=#2,regular polygon, regular polygon sides=3,inner  sep=1.3pt] at (5cm,0) {};
  \end{tikzpicture}%
}

\newcommand*{\tikzline}[1]{%
  \setbox0=\hbox{\strut}%
  \begin{tikzpicture}
    \useasboundingbox (-0.2em,-0.2em) rectangle (1.8em,\ht0);
    \draw[color=#1,solid,line width=0.5pt](0,0) -- (5mm,0);
  \end{tikzpicture}%
}

\newcommand*{\tikzdashedline}[1]{%
  \setbox0=\hbox{\strut}%
  \begin{tikzpicture}
    \useasboundingbox (-0.2em,-0.2em) rectangle (1.8em,\ht0);
    \draw[color=#1,dashed,line width=0.5pt](0,0) -- (5mm,0) {};
  \end{tikzpicture}%
}

\begin{document}

\preprint{XXX/XXXX}

\title{Accounting for inertia effects to access the high-frequency microrheology of viscoelastic fluids }

\author{P. Dom\'{i}nguez-Garc\'{i}a}
\affiliation{Dep. de F\'{i}sica de Materiales, Universidad Nacional de Educaci\'{o}n a Distancia (UNED), Madrid 28040, Spain} 
\email{pdominguez@fisfun.uned.es}

\author{Fr\'{e}d\'{e}ric Cardinaux}
\affiliation{Department of Physics, University of Fribourg, 1700 Fribourg Perolles, Switzerland}
\affiliation{LS Instruments AG, Passage du Cardinal 1, CH-1700 Fribourg, Switzerland}

\author{Elena Bertseva}
\author{L\'{a}szl\'{o}  Forr\'{o}} 
\affiliation{Laboratory of Physics of Complex Matter, Ecole Polytechnique F\'{e}d\'{e}rale de Lausanne (EPFL), 1015 Lausanne, Switzerland}

\author{Frank Scheffold}
\affiliation{Department of Physics, University of Fribourg, 1700 Fribourg Perolles, Switzerland}

\author{Sylvia Jeney}
\affiliation{Laboratory of Physics of Complex Matter, Ecole Polytechnique F\'{e}d\'{e}rale de Lausanne (EPFL), 1015 Lausanne, Switzerland}

\date{\today}

\begin{abstract}
We study the Brownian motion of microbeads 
immersed in water and in 
a viscoelastic wormlike micelles solution by 
optical trapping interferometry and diffusing wave spectroscopy.
Through the mean-square displacement obtained from both techniques,
we deduce the mechanical properties of the fluids at high frequencies by explicitly accounting for
inertia effects of the particle and the surrounding fluid at short time scales. 
For wormlike micelle solutions, we recover 
the 3/4 scaling exponent for the loss modulus over two decades in frequency as predicted by the theory for semiflexible polymers.
\end{abstract}


\maketitle

The quantitative stochastic description of Brownian motion \cite{einstein_motion_1905} 
of spherical micro- and nanobeads in
a complex fluids has laid the foundations for the invention of \emph{tracer microrheology} \cite{mason_optical_1995, waigh_microrheology_2005}, a powerful, 
noninvasive method that allows the measurement of mechanical properties over an extended range of frequencies using all optical instrumentation. 
At very short time-scales, or high frequencies, the stochastic description of Brownian motion fails as pointed out already in the original 
work of Einstein \cite{einstein_motion_1906}. 
At microsecond time scales 
the influence of inertial effects and hydrodynamic memory becomes sizable \cite{franosch_resonances_2011, li_t_brownian_2013}. 
Failing to account for these contributions leads to substantial errors. 
Removing these effects \cite{Liverpool_2005,atakhorrami_short-time_2008} at high frequencies $\omega$ when calculating the complex
modulus, $G^*(\omega)$, may allow one to discern relevant data that is otherwise difficult or impossible to access. 
Elastic moduli at such high frequencies may contain important information about living cells \cite{Fabri_scaling_2001}, 
biopolymers in pharmaceutical applications \cite{hyaluronanOel} or 
fast processes encountered, for example, in ink-jet printing \cite{inkjet}. 

In this work, we demonstrate how to correct for the influence of 
inertia in an actual experiment. We study the Brownian motion of microbeads immersed in water and in a 
viscoelastic wormlike micelle solution by two complementary experimental 
techniques accessing the MHz frequency range: optical trapping microscopy (OTI) and diffusing wave spectroscopy (DWS).
The combined application of these two methods is unique since it covers 
the most relevant approaches to high-frequency microrheology, while other methods, such as particle tracking, 
are limited to frequencies well below 10 kHz \cite{Gardel_microrheology_2003}.
We account for inertia effects quantitatively using two different approaches: 
the self-consistent correction of the mean-square displacement suggested in \cite{Willenbacher_broad_2007} 
and a theoretical expression derived from the recent work of Schieber 
and collaborators \cite{indei_t_competing_2012, cordoba_elimination_2012}.  
We find that these two different quantitative methods provide similar results, as shown by a study of the high-frequency 
scaling of the modulus $G''(\omega) \sim \omega^{\alpha}$ of a viscoelastic wormlike micelle solution. 
Taking into account inertia effects we find an exponent of $\alpha \cong 0.75$, as predicted by the theory for semiflexible polymers \cite{gittes_dynamic_1998}. 
Besides, the proposed methodology allows one to extract parameters relevant to the intrinsic properties of the viscoelastic fluid, 
such as, the bending modulus, the mesh size and the contour length of the molecular components.

In a tracer microrheology experiment, the complex modulus of a bulk material is calculated from 
the measured mean-square displacements MSD $\equiv \left<[\textbf{r}(t)-\textbf{r}(0)]^2\right>$ of microbeads with 
position $\textbf{r}(t)$ \cite{mason_optical_1995,levine_one-_2000}. OTI is the more versatile 
of the two techniques as it acts on individual beads and, therefore, does not 
rely on averaging over the motion of many beads, as is the case for DWS. 
In OTI, the movement of a single microbead is recorded by means of an interferometric position detector \cite{jeney_s_monitoring_2010}. 
The bead is trapped in the center of a sample chamber using optical 
tweezers \cite{ashkin_applications_1980},  applying the lowest optical force 
available, $3 \lesssim k \lesssim 10$ $\mu\text{N}/\text{m}$, where $k$ is the spring constant of the optical restoring force. 
DWS is an extension of dynamic light scattering applied to materials with strong multiple
scattering \cite{DWSoriginal1,DWSoriginal2,mason_particle_1997, Gardel_microrheology_2003} 
and it allows precise measurements of 
the three-dimensional movement of tracer beads in the complex fluid. 
The advantages of DWS are its relative ease of use employing 
standard spectrophotometer cuvettes and the minute-scale measurement times with little calibration and post-processing required. 
DWS also provides access to a large range of experimental parameters and can be applied to viscoelastic fluids and 
solids with low- or high-frequency moduli while covering an extended 
range of frequencies from $\omega \sim 0.1$ to $10^6$ rad/s \cite{Willenbacher_broad_2007}. 
\begin{figure}
\includegraphics[scale=1]{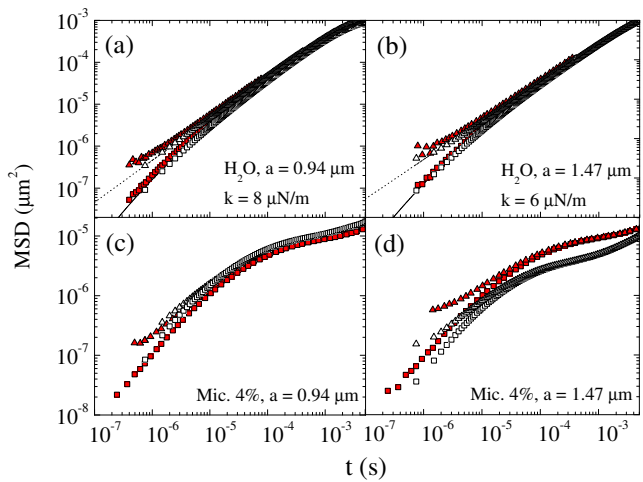}
\caption{\label{fig:fig1}(Color online) One-dimensional particle MSD 
in water and in a 4\% micellar solution for two different bead sizes $a$.  
Experimental data: OTI (\protect\tikzrectangle{black}{white}) 
and DWS (\protect\tikzrectangle{black}{red}). 
Corrected MSDs are shown as triangles: from OTI (\protect\tikztriangle{black}{white}) and
from DWS (\protect\tikztriangle{black}{red}). Theoretical predictions for water: 
Chandrasekhar's result ignoring inertia effects (\protect\tikzdashedline{black}) and Hinch's prediction 
including the influence of inertia (\protect\tikzline{black}).}
\end{figure} 
By directly comparing the results obtained by each of the techniques, we study the Brownian motion of melamine resin microbeads 
 with radii $a=0.94$ or $1.47$ $\mu$m, with density $\rho_p = 1570$ kg/m$^3$ at $T= 21$ ºC. 
 The relatively high refractive index ($n = 1.68$) provides good trapping efficiency in OTI experiments. 
 It has been reported in \cite{valentine_colloid_2004} that the colloid 
 surface chemistry of the microbeads can affect the results obtained by microrheology. 
 Such effects are particularly relevant for protein specific bonding in biomaterials. 
 In our case, however, the resin beads are chemically non-active and therefore 
 we expect such effects to be negligible. 
 The sample under study is an aqueous solution at 4 wt \% of surfactant cetylpyridinium 
 chloride (CPy$^+$Cl$^-$) and sodium salicylate (Na$^+$Sal$^-$) that self-assemble into a wormlike micelle 
 solution \cite{CatesCandau90}. At sufficiently high frequencies the cylindrical 
 micelles are expected to behave as a solution of semiflexible polymers in 
 the semi-dilute regime formed by entangled micelles \cite{buchanan_high-frequency_2005}. 
 We measure the steady-state viscosity using a Rheometer MCR502 (Anton
Paar, Austria) at $21$ ºC, which yields to $\eta_0 = 360$ $\text{mPa}\cdot\text{s}$.
 In the OTI experiments, a double flow chamber is used: One of the chambers contains water 
 and beads at very low concentration, while the other one encloses 
 the viscoelastic fluid with the same type of beads. 
 After aligning the optical trapping light path to the position detector and adjusting signal amplification, 
 a calibration experiment is performed in the pure water solution. 
 The measurement in the viscoelastic solution is done right afterwards.
 This procedure allows using the known properties of a Newtonian fluid \cite{grimm_high-resolution_2012} to extract the 
 volts-to-meter conversion factor $\beta$, which is then used to calculate the MSD in the 
viscoelastic medium \footnote{Calibration values for Fig. 1 are $\beta=10.3\,\mu$m/V for $a=0.94\,\mu$m and $\beta=19.9\,\mu$m/V for  $a=1.47\,\mu$m.}. 
For the DWS experiments, we are using a bead 
 concentration of 2 vol. $\%$. Samples are loaded into standard rectangular spectrometer cuvettes with a path length $L=2$ mm  or $L= 5$ mm and a 
width of 10 mm. Echo two-cell DWS experiments in transmission geometry are performed as 
described in Ref. \cite{zcs06}. 
The two experimental techniques provide similar MSDs for the bead motion in water, as shown in Figs. \ref{fig:fig1}(a) and \ref{fig:fig1}(b). 
However, for the micelle solution,
some differences are observed [Figs. \ref{fig:fig1}(c) and \ref{fig:fig1}(d)], in particular for the larger bead size. 
Similar global shifts of the MSD have been observed previously \cite{Cardinaux2002,Oelschlaeger_linear_2008}. They have been 
attributed to hydrodynamic effects at the fluid-particle interface 
and local perturbations of the equilibrium configuration of the complex fluid \cite{chen_rheological_2003}. 
The optical force in a micelle solution might trap some of the polymeric structure, hindering the Brownian motion and, thereby, slightly incrementing the apparent 
measured viscosity of the fluid. DWS in turn is sensitive to depletion of the surfactant solution around the beads and a possible 
onset of depletion-induced bead attraction which can result in enhanced motion and larger MSD values \cite{Oelschlaeger_linear_2008}.
To observe the effects of inertia and hydrodynamics in the MSDs, 
we compare in Fig. \ref{fig:fig1}(a) and \ref{fig:fig1}(b) the results obtained for bead motion in pure water with the Chandrasekhar expression 
for a classic Newtonian fluid \cite{chandrasekhar_stochastic_1943}, given by
$\left<\Delta r^2(t)\right> = [1-\exp(-kt/\gamma)]\,2k_B T/k$ where $\gamma = 6\pi\eta a$. For the case of DWS, there is no 
optical trap and the classical Stokes-Einstein result $\left<\Delta r^2(t)\right> = (k_BT/3 \pi\eta a)\,t$ is recovered in the limit $k\to0$. 
In both cases, substantial deviations due to 
inertial effects are observed at times shorter than $t=10^{-4}$ s. For water, the inertia effects in the MSD 
can be reproduced quantitatively using the classical result according 
to Hinch \cite{Hinch75} [Fig. \ref{fig:fig1}(a) and \ref{fig:fig1}(b)]. 

The standard formalism to convert the measured MSD to the complex elastic modulus $G^*(\omega)$ is the Mason-Weitz (MW) approach based on 
the generalized Stokes-Einstein relation (GSER) \cite{mason_particle_1997} concurrent with Mason's approximation \cite{mason_estimating_2000} to obtain:
\begin{equation}
Z^*(\omega)  = \frac{k_B T}{i \omega \pi a \overline{\left<\Delta r^2(\omega)\right>}},\label{GSER1}
\end{equation}
where $\overline{\left<\Delta r^2(\omega)\right>}$ is the one-side
Fourier transform of the MSD. In the absence of inertia effects, $G^*(\omega) \equiv Z^*(\omega)$ for 
equilibrium Brownian motion in a homogeneous viscoelastic fluid \cite{levine_one-_2000}.
An alternative methodology to connect the MSDs with the 
rheological properties of the fluid was reported 
by Evans \textit{et al} \cite{evans_direct_2009}.
Both methods \footnote{See Supplemental Material for a comparison of the two methodologies using our data.} have common limitations, such as the omission of optical or external 
forces, the neglect particle and fluid inertia and the non-consideration of active and heterogeneous materials \cite{squires_fluid_2010}. 
\begin{figure}
\begin{center}
\includegraphics[scale=1]{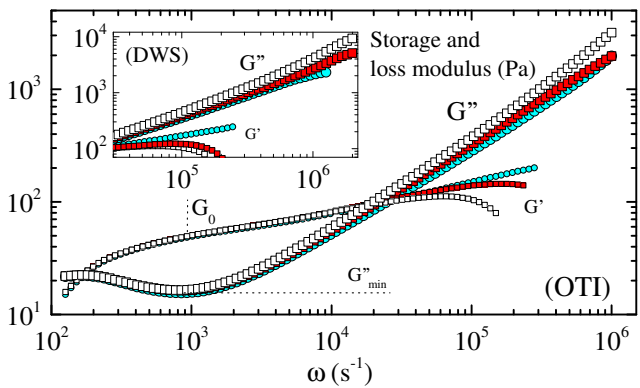}
\caption{\label{fig:fig2} (Color online) Comparison of the microrheology results using OTI and DWS (Inset) for one bead size $a=0.94$ $\mu$m in 
a viscoelastic micelle solution with and without taking into account the effects of inertia. 
Data without inertia correction (\protect\tikzrectangle{black}{white}), corrected using Eq. (\ref{indei2}) (\protect\tikzrectangle{black}{red}) and
using the corrected-MSDs shown in Fig. \ref{fig:fig1} (\protect\tikzbullet{black}{cyan}).
} 
\end{center}
\end{figure}
The inertia effects appearing in the
power-spectral density (PSD) of probe particles 
using optical tweezers were studied in Refs. \cite{Peterman2003, Berg2004}, 
while a systematic approach to account for inertia effects 
in the MSD obtained from passive bead microrheology 
has been described in Ref. \cite{Willenbacher_broad_2007}. 
The principal idea in the latter work is to define an 
effective viscosity of the medium by $\eta_{\text{eff}}(t)  = [k_BT/3 \pi \left<\Delta r^2(t)\right> a]\, t$ 
and then calculate, for each measured time point $t$, the correction 
factor $f(t,\eta_{\text{eff}}(t))\ge 1$ for the MSD using the known theoretical result 
for a Newtonian fluid \cite{Hinch75}. The procedure is iterated numerically starting with 
the measured MSD. The MSD obtained after correction is used for further processing in Eq. (\ref{GSER1}). 
The same strategy has been applied in Ref. \cite{mizuno_active_2008} using  
the PSD \footnote{In that work, 
the Fourier transform was carried out first on the measured time series 
of bead positions to obtain the PSD. The inertia correction is then executed 
on the PSD using the known expression for the frequency dependent particle mobility in a 
simple liquid using an effective complex viscosity of the medium derived from the 
experimental data.}.
A more fundamental theoretical treatment has been suggested by 
Felderhof \cite{Felderhof2009} and more recently by Indei \textit{et al.} \cite{indei_t_competing_2012} 
and C\'{o}rdoba \textit{et al.} \cite{indei_t_competing_2012,cordoba_elimination_2012}. The 
latter work conveniently provides an analytical expression relating the actual 
complex modulus and the bead mean-square displacement, 
assuming that the medium is incompressible \footnote{Equation (37) of Ref. \cite{indei_t_competing_2012} is the following 
$G^*(\omega) = \frac{k_B T}{i \omega \pi a \overline{\textrm{MSD}}}+\frac{m^*\omega^2}{6\pi a}
+\frac{a^2 \omega^2}{2}\times \left[ \sqrt{\rho^2+\frac{2\rho}{3\pi a^3}\left(\frac{6k_B T}{(i\omega^3 \overline{\textrm{MSD}})}-m^*\right)}-\rho \right]$,  
which can be related to Eq. (\ref{GSER1}) since the term in the square root is 
$6 k_B T/(i\omega^3 \overline{\textrm{MSD}})=-(6 \pi a/\omega^2)\,Z^*(\omega)$}: 
\begin{align}
G^*(\omega) &= Z^*(\omega) +\frac{m^*\omega^2}{6\pi a}+\frac{a^2 \omega^2}{2}\notag \\
&\times \left[ \sqrt{\rho^2-\frac{2\rho}{3\pi a^3}\left(\frac{6 \pi a}{\omega^2}\,Z^*(\omega) +m^*\right)}-\rho \right],\label{indei2}
\end{align} 
where $m^{\ast}=m_{\text{particle}}+ 2 \pi \rho a^3/3$ is the effective mass of the particle.
By calculating $Z^*(\omega)$ and then using Eq. (\ref{indei2}), we obtain the 
complex moduli without the influence of inertia. 
It is important to note that other possible corrections to the high-frequency bead motion, 
for example due deviations from the GSER at $\omega \ge 10^6$ Hz, are not included
by these corrections \cite{levine_one-_2000}. 

We apply these correction protocols to the experimental MSDs displayed in 
Fig. \ref{fig:fig1} to obtain the complex modulus 
for the micelle solution. In Fig. \ref{fig:fig2} 
we show the results derived from DWS and OTI using a bead size $a=0.94$ $\mu$m \footnote{An equivalent 
figure for $a = 1.47$ $\mu$m is included in the Supplemental Material} for OTI and DWS, with and without corrections. 
The inertia-corrected results for $G''(\omega)$ calculated using both methods yield to similar results. 
These results can be compared to the 
the expected behavior at higher frequencies for a solution of semiflexible 
polymers, which, according to Gittes and MacKintosh \cite{gittes_dynamic_1998}, is given by:
\begin{equation}
G^{\ast}_{\textrm{GMK}}(\omega) = i\omega\eta_s + \frac{1}{15}\rho_m \kappa \l_p \left(\frac{-2i\xi}{\kappa}\right)^{3/4} \,\omega^{3/4}\label{ggittes}
\end{equation}
\begin{figure}
\begin{center}
\includegraphics[scale=1]{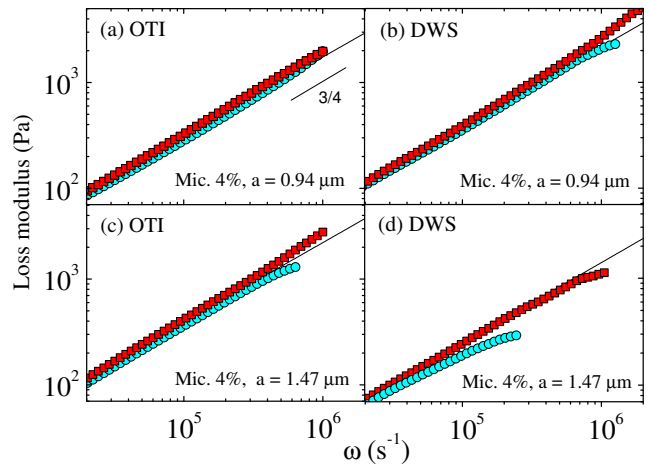}
\end{center}
\caption{\label{fig:fig3}(Color online)
High-frequency loss modulus, $G''(\omega)-\omega \eta_s$ compared to theoretical predictions.   
Inertia-corrected data using Eq. (\ref{indei2}) (\protect\tikzrectangle{black}{red}) and
using the corrected-MSDs shown in Fig. \ref{fig:fig1} (\protect\tikzbullet{black}{cyan}).
(a) OTI, $a=0.94$ $\mu$m.
(b) DWS, $a=0.94$ $\mu$m. (c) OTI, $a=1.47$ $\mu$m.
(d) DWS, $a=1.47$ $\mu$m. 
Black line (\protect\tikzline{black}) is $G''_{\text{GMK}}(\omega)-\omega \eta_s$ 
evaluated using Eq. (\ref{ggittes}) and data from Table \ref{tab1}.
} 
\end{figure}
\begin{table*}
\caption{\label{tab1} Characteristic magnitudes for the wormlike micelle solution obtained from OTI
and DWS using Eq. (\ref{ggittes}). 
For all data: $d_{\textrm{mic}} = 2.5$ nm and area density $\rho_m = 8.2 \times 10^{15}$ m$^{-2}$.
}
\begin{ruledtabular}
\begin{tabular}{ccccc|cccccccc}
$a$ ($\mu$m) &Tech. &$\omega_0$ (s$^{-1}$ $\times 10^4$) &$G_0$ (Pa) &$G''_{\textrm{min}}$ (Pa) &$l_p$ (nm) &$\kappa$ (Jm $\times 10^{-29}$) &$\xi$ (nm) &$\zeta$ (Ns/m$^2 \times 10^{-3}$) &$l_e$ (nm) &$L$ (nm)\\
\hline
$0.94$ &OTI &$1.7 \pm 0.1$ &$50 \pm 5$ &$15 \pm 2$ &$31.0 \pm 0.6$ &$12.6 \pm 0.3$ &$43 \pm 1$ &$5.4 \pm 0.2$ &$54 \pm 4$ &$180 \pm 50$\\
$0.94$ &DWS &$1.0 \pm 0.1$ &$50 \pm 5$ &$15 \pm 2$ &$37 \pm 1$ &$15.0 \pm 0.5$ &$43 \pm 1$ &$5.4 \pm 0.2$ &$48 \pm 4$ &$160 \pm 50$\\
$1.47$ &OTI &$1.0 \pm 0.1$ &$60 \pm 6$ &$30 \pm 3$ &$37 \pm 1$ &$15.0 \pm 0.5$ &$41 \pm 1$ &$5.2 \pm 0.2$ &$43 \pm 3$ &$90 \pm 20$\\
$1.47$ &DWS &$3.0 \pm 0.1$ &$60 \pm 6$ &$30 \pm 3$ &$25.7 \pm 0.3$ &$10.4 \pm 0.1$ &$41 \pm 1$ &$5.2 \pm 0.2$ &$69 \pm 3$ &$140 \pm 40$\\
 \end{tabular}
 \end{ruledtabular}
 \end{table*}
Then, it is expected that both moduli, loss and storage, follow the $G^*\sim \omega^{3/4}$ behavior at high frequencies. However, 
at high frequencies the viscous response of the material dominates the elastic response 
by about an order of magnitude. This makes it very difficult to extract meaningful 
information about $G'$ from microrheology. We thus restrict our discussion to the comparison of Eq. (\ref{indei2})
using only the experimental loss modulus, $G''$, for CpyCl/NaSal to study in more detail the accuracy of inertia corrections at high frequencies 
\footnote{Using the GSER at high frequencies
can introduce non-expected effects. See Supplemental Material document for extra comments about this point.}.
Next, we compare Eq. (\ref{ggittes}) using only the experimental loss modulus for CpyCl/NaSal to study 
in more detail the accuracy of inertia corrections at high
frequencies. To evaluate Eq. (\ref{ggittes}), we use the standard theory of polymers \cite{doi_theory_1986}  
and the experimental $G_0$, $G''_{\textrm{min}}$, and $\omega_0$ as input parameters.
The quantity $G_0$ is the value of $G'$ at which $G''$ has a local minimum, denoted as $G''_{\textrm{min}}$. Both characteristic values 
can be easily obtained from Fig. \ref{fig:fig2}. The third input parameter is $\omega_0$, the 
crossover frequency where the exponent $\alpha$ of the power-law behavior $G''\sim \omega^\alpha$
changes from the Rouse-Zimm behavior 
to the expected 3/4 at higher frequencies \footnote{For DWS experiments we use $G_0$ and $G''_{\textrm{min}}$ from OTI. We select $\omega_0$
for the experimental $G^*$ to match $G^*_{\textrm{GMK}}$, 
obtaining $\omega_0 \sim 10^4\, \textrm{Hz}$, consistent with literature values}. 
The persistence length of the polymer-like micelles is then linked to $\omega_0$ by  
$l_p = (kT/8\eta_s\omega_0)^{1/3}$. The micelles diameter is estimated as $d_{\textrm{mic}} = 2.5$ nm \cite{Willenbacher_broad_2007} and the area 
density is $\rho_m = \phi/[(\pi/4) d^2_{\textrm{mic}}] = 8.2 \times 10^{15}$ m$^{-2}$, where $\phi$ is the volumetric concentration \cite{Oelschlaeger_linear_2008}. 
Then, we calculate the bending modulus, $\kappa=kT \,l_p$, the mesh size, $\xi = (kT/G_0)^{1/3}$, 
the lateral drag coefficient, $\zeta=4\pi\eta_s/\ln(0.6\lambda/d_{\textrm{mic}})$
where $\lambda = \xi$, the contour length between two entanglements, $l_e=\xi^{5/3}/lp^{2/3}$, and the contour length of the micelles, $L = l_e \,G_0/G''_{\textrm{min}}$.
The results of the calculations for these quantities are summarized in Table \ref{tab1}. 
The different experimental methods (DWS and OTI) as well as the two bead sizes give similar results. 
The averaged persistence length, $l_p\sim 33 \pm 6$ nm, agrees with previous mechanical measurements for this kind of 
surfactant solution (29 nm) \cite{Willenbacher_broad_2007}, but the averaged mesh size, 
$\xi \sim 42$ nm, is slightly lower (52 nm). To quantify the inertia corrections for the viscoelastic fluid at frequencies $10^4$-$10^6$ Hz, 
we plot in Fig. \ref{fig:fig3} the loss modulus obtained from applying the two different correction procedures. 
The results quantitatively agree with theory, except for some deviations due to the limited accuracy of the experimental methods at higher frequencies.
Equation (\ref{GSER1}) applied to the inertia-corrected MSDs and to the Indei-Schieber method [Eq. (\ref{indei2})] yield to similar
results, matching the theoretical curves and the $\omega^{3/4}$ behavior, especially when using OTI and small beads. 
Differences between both methods arise when using DWS and bigger beads, where Eq. (\ref{indei2}) provides better results. 
\newline \indent In conclusion, our experimental data and analysis demonstrate that, 
when properly accounting for inertia effects, passive microrheology easily provides access to the mechanical properties
of viscoelastic fluids up to frequencies in the MHz scale. The correction procedure 
is simplified by the availability of an analytical expression that can be applied 
straightforwardly to the experimental data. This procedure extends the range of application of 
microrheology up to two orders of magnitude, and therefore will enable 
new experiments in a regime which is fairly inaccessible to traditional mechanical rheometry. 
\newline \indent We acknowledge B. U. Felderhof and T. Franosch for helpful discussions.
P.D.G acknowledges M.E.C.D. for financial aid by Plan Nacional I+D+i 2008-2011 and project FIS2013-47350-C5-5-R. P.D.G.
and S.J. acknowledge support from the NCCR-Nano (project 1.4). F.S. acknowledges 
financial support by the Swiss National Science Foundation under Grants No. 132736 and No. 149867.

\end{document}